\documentstyle [preprint,revtex]{aps}

\begin{document}
\begin{flushright}
VAND-TH-92-3\\
March 1992\\
(Revised)\\
\end{flushright}
\begin{center}
\LARGE
Gauge Field Back-reaction on a Black Hole
\end{center}
\normalsize
\bigskip
\begin{center}
{\large David Hochberg} and {\large Thomas W. Kephart}\\
{\sl Department of Physics and Astronomy, Vanderbilt University\\
Nashville, TN 37235, USA}\\
\end{center}
\vspace{2cm}
\centerline{ABSTRACT}
The order $\hbar$ fluctuations of gauge fields in the vicinity
of a blackhole can create a repulsive
antigravity region extending out
beyond the renormalized Schwarzschild horizon. If the strength of this
repulsive force increases as higher orders in the back-reaction
are included, the formation of a wormhole-like object
could occur.
\vfill\eject
\noindent
{\bf 1. Introduction}

The study of wormholes, or handles in the spacetime topology, has been
an active research area in recent years, with the bulk of the effort
going into understanding the behavior of topology change in Euclidean
space and the putative role these instantons might have for solving
the cosmological constant problem, spawning baby Universes, and related
concerns [1]. In physical (i.e., Minkowski-signature)
spacetime by contrast, wormholes have
attracted less attention, probably due to the lack (until recently)
of concrete and physically meaningful examples.
Indeed, the only explicit
example for over seventy years has been provided by the well-documented
Schwarzschild wormhole [2], which is at best a theoretical curiosity.
Not only does the wormhole throat coincide with the blackhole event
horizon, but the ``bridge" connecting the two asymptotically flat
spacetimes snaps into two disconnected pieces so rapidly as to prevent
any signal (even light) from traversing the throat [3]. However, a
recent analysis of static isotropic solutions of Einstein's equation
revealed these solutions contain bona-fide wormholes provided the
throat is threaded with stress-energy violating the weak energy condition
(WEC) [4]. These wormholes are free from the problems plaguing
the Schwarzschild case.
If the requisite stress-energy exists, then one should attempt to
pin down and expose the physical traits belonging to this class
of wormholes and to clarify the scenarios leading to their
existence. The knowledge derived from this will lead to
a better understanding of the
kind of nontrivial topologies allowed
by the gravitational field equations
and motivates exploring the cosmological consequences
these wormholes are bound to have in the early Universe.

A hitch is encountered at the purely classical level however, as no
classical stress tensor is known to violate the WEC.
But the interplay of gravity and quantized matter can lead to the
type of stress-energy relevant for maintaining a wormhole structure
since quantum fields in curved backgrounds tend to
develop localized negative
energy densities [5].
Since the production of negative energy densities is a generic
consequence of the gravitational interaction with matter [6], this
lends credence to the picture of
the early Universe as a spacetime
containing naturally occuring wormholes.
The task remains to find an explicit wormhole solution arising
from a physically realized matter source.
As a step in this direction, in this paper we derive
a semiclassical solution of Einstein's equation coming from
the influence of quantized gauge fields on the
classical background geometry of a Schwarzschild
blackhole. This means we are to solve the back-reaction problem
$$G_{\mu \nu}({\hat g}, \delta g) = 8\pi <T_{\mu \nu}>,\eqno(1.1)$$
where $<T_{\mu \nu}>$ represents a renormalized quantum stress tensor
for matter in a Schwarzschild background (with metric ${\hat g}$), and
$\delta g$ is the perturbation caused by the source term. The exact
1-loop renormalized quantum stress tensor has been calculated for
$U(1)$ fields [7], and we take this as our source term. We also
consider the back-reaction of non-abelian gauge fields in the
limit where the gauge self-coupling is small. In that limit, we
can take over the results in [7] by noting that the abelian
and non-abelian stress tensors are related by
$$<T_{\mu \nu}>_{G} \,\approx \, C_2(G) <T_{\mu \nu}>_{U(1)},\eqno(1.2)$$
where $C_2(G)$ is the second Casimir invariant (for the group $G$)
evaluated in the adjoint representation.
If $C_2(G) \stackrel{>}{\sim} 100$, (e.g., the string-inspired
unification group $E_8 \times E_8$ has $C_2 = 120$) the total
effective mass seen by a static observer located within a few times
the renormalized Schwarzschild radius
can be {\it negative}, and hence she will feel
a repulsive gravitational force in that region. As discussed in
[4], this repulsion or ``defocussing" is a necessary condition for
the existence of a traversable wormhole.

\noindent
{\bf 2. The Solution of the Back-reaction}

A concise discussion of the back-reaction problem along with details
useful for calculating on Schwarzschild backgrounds can be found in
[8]. There, the Eddington-Finkelstein coordinates were shown to be
useful for reducing the solution of (1.1) to two
elementary integrations. In these
coordinates, the
(background) Schwarzschild metric takes the form
$$ds^2 = - (1 - 2M/r)dv^2 + 2dv\,d{\tilde r} + {\tilde r}^2 d\Omega^2,
\eqno(2.1)$$
where $v = v(t,r) = t + r + 2M\ln(r/2M -1)$ and ${\tilde r} = {\tilde r}
(r) = r$.
To investigate the changes induced in this geometry by the
presence of matter or other fields, we require a suitable
ansatz for the complete metric $g = {\hat g} + \delta g$. For
radially symmetric and static source terms the most general
metric is given by
$$ds^2 = -e^{2\psi}(1 - 2m/r)\,dv^2 + 2e^{\psi}\,dv\,d{\tilde r} +
{\tilde r}^2\,d\Omega^2,\eqno(2.2)$$
where $\psi$ and $m$ are two radially dependent functions to be
determined by solving (1.1). This reduces to (2.1) in the limit when
the source term is shut off.

The Universe abounds in gauge bosons, and it is therefore of interest
to consider in what way these particles can alter the spatial
geometry surrounding compact massive objects. This question can now
be investigated by making use of a recent calculation of the
renormalized electromagnetic stress
tensor in a Schwarzschild background.
This tensor can be decomposed into the
two pieces
$$<T^{\nu}_{\mu}> = <T^{\nu}_{\mu}>_{analytic} + {1 \over {\pi^2 (4M)^4}}
 \,\Delta^{\nu}_{\mu},\eqno(2.3)$$
where $<T^{\nu}_{\mu}>_{analytic}$ is given in closed form and
$\Delta^{\nu}_{\mu}$ comes from a numerical evaluation of a mode sum [7].
Since this latter contribution is very small in comparison to the analytic
part, we will not include it in our calculations. The analytic part of the
tensor is a simple sixth-degree polynomial
given by (we drop the angular brackets) [7]
$$T^{t}_{t} = {{\pi^2 \hbar} \over {45 (8\pi M)^4}} (-3 - 6w -9w^2 -12w^3
+ 315 w^4 -78w^5 + 249w^6),\eqno(2.4)$$
$$T^{r}_{r} = {{\pi^2 \hbar} \over {45 (8\pi M)^4}} (1 + 2w + 3w^2 -76w^3
+ 295w^4 -54w^5 + 285w^6),\eqno(2.5)$$
and
$$T^{\theta}_{\theta} = T^{\phi}_{\phi} = {{\pi^2 \hbar} \over {45 (8\pi M)
^4 }} ( 1 + 2w + 3w^2 + 44w^3 - 305w^4 + 66w^5 -579w^6),\eqno(2.6)$$
where $w \equiv 2M/r$. It is straightforward to show this tensor is
conserved with respect to the background: ${\hat \nabla}_{\mu}
T^{\mu}_{\nu} = 0$. It is crucial this hold in order that the
Bianchi identity not be violated.
Moreover, one sees by inspection that the renormalized stress tensor
is asymptotically constant. As $r \rightarrow \infty$,
$T^{\mu}_{\nu} \rightarrow {{\pi^2 \hbar} \over 45} T^4_{H}\,
diag(-3,1,1,1)$, which is precisely the form of a stress tensor for
a thermal ensemble of radiation at a temperature $T_{H} = 1/(8\pi M)$.
For the Hartle-Hawking thermal state, which represents a blackhole
in thermal equilibrium with radiation, the polynomial expressions
for $T^{\mu}_{\nu}$ reduce at infinity to that of thermal radiation
in flat spacetime. This must be the case in order that an observer
at infinity measure the correct value of the Hawking temperature,
 $T_{H}$. Thus, the asymptotic constancy of the stress tensor
is a physical requirement.
The mass-energy density as seen
by a static observer is $\rho = -T^{t}_{t}$,
and is negative from $r = 2M$ out to $r \approx 5.14M$, as shown
in Fig. 1. As we will see below, this ``pocket" of negative energy
screens the positive mass $M$ of the singularity from observers located
in the neighborhood of the horizon.
The existence of this negative energy density implies the violation
of both the weak and strong energy conditions. Moreover, the radial
tension $(-T^{r}_{r})$ exceeds the energy density in the range
$2M < r \stackrel{<}{\sim} 3.2M$, which results in
the breakdown of the dominant
energy condition. We mention these points as (2.3) provides an explicit
example of a physical stress tensor which therefore violates all
the fundamental energy assumptions of
classical general relativity [9].

When cast in terms of the coordinates (2.1),
these components are given by $T^{v}_{v} = T^{t}_{t}$,
$T^{\tilde r}_{\tilde r} = T^{r}_{r}$ and
$$T^{v}_{\tilde r} = (1 - 2M/r)^{-1}\,[T^{r}_{r} - T^{t}_{t}],\eqno(2.7)$$
and $T^{\theta}_{\theta} = T^{\phi}_{\phi}$ are unchanged. The corresponding
components of the Einstein tensor which derive from (2.2) are
$$G^{v}_{v} = {-2 \over {r^2}} \,{{dm} \over {dr}},\eqno(2.8)$$
$$G^{v}_{\tilde r} = {2 \over r} e^{-\psi} {{d\psi} \over {dr}},
\eqno(2.9)$$
and
$$G^{\tilde r}_{\tilde r} = {2 \over {r^2}}\,
\left[ r(1 - 2m/r) {{d\psi} \over {dr}} - {{dm} \over {dr}}
\right].\eqno(2.10)$$
Following [8], we introduce the expansion parameter $\epsilon = \hbar/M^2$.
(in units where $c=G=1$, but $\hbar \neq 1$).
Since the stress tensor (2.3) is known to one-loop, it is $O(\hbar)$, and
consistency demands we solve for the metric functions only to the same
order. We therefore expand
$$e^{\psi} = 1 + \epsilon A(r), \;\; m = M[1 + \epsilon B(r)],\eqno(2.11)$$
which when inserted into (1.1) yield the linearized Einstein equations
$$-{{2M\epsilon} \over {r^2}}\, {{dB} \over {dr}} = 8\pi T^{t}_{t},
\eqno(2.12)$$
$${{2\epsilon} \over r}\, {{dA} \over {dr}} = 8\pi (1 - 2M/r)^{-1}\,
(T^{r}_{r} - T^{t}_{t}),\eqno(2.13)$$
and
$${{2\epsilon} \over {r^2}} \left[ (r - 2M) {{dA} \over {dr}} - M
{{dB} \over {dr}} \right] = 8\pi T^{r}_{r}.\eqno(2.14)$$
The substitution of (2.12) and (2.13) into (2.14)
reproduces the identity (2.7), while
the $\theta$ and $\phi$ components of (1.1) are automatically satisfied
since the corresponding constraints ${\hat \nabla}_{\mu} T^{\mu}_{\theta} =
{\hat \nabla}_{\mu} T^{\mu}_{\phi}$ are vacuous. Thus, the backreaction
problem reduces to solving the two equations (2.12) and (2.13).
Using (2.4) and (2.5) we integrate (2.12) and (2.13) to find that
$$KB(r) = 2\left[{83 \over 3}w^3 - 13w^2 +105w - 4ln(w) + {3 \over w}
+ {1 \over {w^2}} + {1 \over {3w^3}} \right] - 248 +KB(2M),\eqno(2.15)$$
and
$$KA(r) = {4 \over 3}\left[ 3w^3 + {15 \over 2}w^2 + 10w - 6ln(w)
 + {3 \over w} + {1 \over {2w^2}} \right] - 32 + KA(2M),\eqno(2.16)$$
where $K = 3840\pi$.
The gravitational redshift is given by $A(r)$, and the constant of
integration $A(2M)$ may be fixed after imposing boundary conditions
at the radius governing the domain of
validity of the perturbation (for a detailed discussion of boundary
conditions, we refer the reader to Sections VII - IX of Reference [8]).
However, the remaining discussion is independent of this constant, and
we do not need to specify it.
The function $B(r)$ plays the role of a mass-energy function
for the gauge fields and is responsible
for determining the shape of the spatial geometry encoded in the
metric (2.2). As discussed in [8], the integration constant $B(2M)$
renormalizes the bare mass $M$ via
$M \rightarrow M_{ren} = M[1 + \epsilon B(2M)],$
and can thus be absorbed into the mass parameter. We shall do so
with the understanding that wherever $M$ appears, it now stands
for the ``dressed" blackhole mass $M_{ren}$,
and the
constant plays no further
role in our calculations.
The resulting total effective mass-function is
$$m_{eff}(r) \equiv M[1 + \epsilon \mu(r)],\eqno(2.17)$$  where
$\mu(r) \equiv  B(r) - B(2M)$ is plotted in Fig. 2. This shows
how the energy density $-T^{t}_{t}$ leads to a negative bosonic mass
density in the vicinity of the blackhole horizon.
This density vanishes
at that horizon,  achieves its most negative value $K\mu(r) \approx -120$
at $r \approx 5.2M$ and passes through zero
at $r \approx 10.8M$, after which
it remains positive definite with the asymptotic behavior
$\sim (r/M)^3$. To reveal the nature of the modified geometry
induced by (2.3),
we transform (2.2) back to diagonal form with coordinates
$(t,r,\theta,\phi)$ by taking (where $m$ stands for $m_{eff}$ from here on)
$${{\partial v} \over {\partial t}} = 1 \;\qquad {\rm and}\; \qquad
{{\partial v} \over {\partial r}}
 = e^{-\psi}(1 - 2m/r)^{-1},\eqno(2.18)$$
which yields the metric
$$ds^2 = -e^{2\psi}(1 - {{2m} \over r})dt^2 + (1 - {{2m} \over r})^{-1}
dr^2 + r^2 d\Omega^2.\eqno(2.19)$$

With the solution (2.19) at hand, we now turn to examine some of its
qualitative features. A convenient way to visualize the solution is
by ``lifting" out the spatial geometry at a fixed time, $t = {\rm
constant}$, and embedding it in a higher dimensional space. The
geometry is spherically symmetric, so without loss of generality,
we confine attention to the equatorial slice $\theta = \pi/2$. The
line element of this slice is given by
$$d{\hat s}^2 = (1 - {{2m(r)} \over r})^{-1}\,dr^2 + r^2\, d\phi^2,
\eqno(2.20)$$
and can be embedded in a three-dimensional embedding space. Typically,
 a single Euclidean space would suffice [4,10], but if the slice
represents a space of negative curvature, one must use a flat
Minkowski space (for example, the 3-geometry of Friedmann
Universes of negative spatial curvature embeds in a 4-dimensional
Minkowski space [10]). In either case, the line element of the (flat)
embedding space is
$$dS^2 = \pm dz^2 + dr^2 + r^2\, d\phi^2,\eqno(2.21)$$
when written in terms of cylindrical coordinates $(z,r,\phi)$.
Identifying the $(r,\phi)$ coordinates of (2.21) and (2.20)
yields the equation for the embedding function
$$ \pm \left( {{dz} \over {dr}} \right)^2 =
 \left( 1 - {{2m(r)} \over r} \right)^{-1} - 1,\eqno(2.22)$$
where $z = z(r)$ decribes a surface of revolution (in the angle
$ \phi$) about the $z$-axis. For example, in the limit
$m \rightarrow M$ (i.e., $\epsilon \rightarrow 0$), the left-hand
side of (2.22) is positive definite (for $r > 2M$), hence
$z$
embeds completely within a Euclidean space. This
reproduces the
standard picture of the Schwarzschild solution with the
derivative diverging at the horizon $r = 2M$ and going to zero
as $r \rightarrow \infty$ [10]. The square root must be taken in
solving for $z$, and both branches meet up at the
Schwarzschild horizon, indicating
the presence of a (non-traversable) wormhole with a throat radius
equal to the horizon radius, as expected.
However, Eq.(2.17) with $\mu \rightarrow C_2\, \mu$ and an
inspection of Fig. 2 indicates
that the combined blackhole-plus-boson mass function $m(r)$ can
become negative before reaching the Schwarzschild horizon.
To embed this region in three dimensions one can
use a Minkowski embedding space (i.e., choose the minus sign in
(2.21)).
Outside that region, the
solution reveals positive definite curvature, and can again be
embedded in a Euclidean space. To embed all of $z$ at once
therefore requires use of a complex embedding space. A somewhat
simpler picture containing the same information is had
by plotting instead $(dz/dr)^2$, so that all regions can be displayed
in a single graph. This is done in Fig. 3, which shows a typical curve
for a solution with a repulsive region. For comparison, we have included
the curve for the Schwarzschild ($\epsilon = 0$) case. In the former
case, the dotted segment represents the antigravity domain.
There are
four length scales associated with this curve. Namely, the
(renormalized) horizon at $r = 2M$, the two points where the total
mass crosses zero
at $r_1$ and $r_2$, and a cutoff radius $r_{dom}$
which serves to define the domain of validity of the calculation
(see below).
Although the repulsive contribution from $\mu(r)$ is too
feeble to overcome $M$ for a single U(1) boson, the effect becomes
pronounced for large numbers of extra abelian gauge fields or for
large non-abelian groups in the limit when the gauge self-coupling
is weak.
In that limit, for any gauge group $G$, we have
$<T_{\mu \nu}>_{G} \, \approx C_2(G) <T_{\mu \nu}>$, where
$C_2(G)$ is the
second Casimir invariant for $G$ in the adjoint representation,
and multiplies the abelian tensor given in (2.4-2.6). The
corresponding embedding function obtains simply by replacing
$\mu  \rightarrow C_2(G) \mu$ in (2.22). We have calculated
the extent of this repulsive domain for a number of gauge groups and
summarize our results in Table I. To stay within the bounds of perturbation
theory, $\epsilon \stackrel{<}{\sim}  1$,
and this implies a minimum $C_2$ for which
the repulsion disappears, namely $C_2 \sim 101$. Thus, the smallest
unitary, orthogonal or abelian groups which can lead to repulsion are
approximately
SU(50), SO(50) and
$[U(1)]^{100}$; the corresponding $r_1$ and $r_2$
are given in Table I. We point out that string-inspired models
containing large numbers of extra abelian gauge fields are not
entirely out of the question. Indeed, both $E_8 \times [U(1)]^{248}$
($C_2 = 308$) and $[U(1)]^{496}$ yield anomaly-free superstring
theories (in ten-dimensions) and
can lead to four-dimensional gauge theories that
contain more than enough extra gauge fields
to generate a repulsive domain [11]. Another anomaly-free
group relevant for string-inspired model building is $E_8 \times E_8$,
with a $C_2 = 120$ (Table I). For a given $C_2$, there is a minimum
$\epsilon$ which leads to a repulsive region; these are given in
Table I.

To put these calculations in perspective, we should determine the
radial
domain of validity of the metric perturbations leading to the
solutions listed in Table I.
That our solution must be truncated at some value of the radius
can be understood as follows. First, recall the stress tensor
is asymptotically constant, so it is clear that the corrected
metric computed from (1.1)
cannot be asymptotically flat. However, our calculation
yields a solution of the back-reaction (1.1) only to lowest order
in $\hbar$, and must therefore be regarded as a strictly perturbative
result. The ``complete" solution of (1.1) can in principle be
obtained through an iterative approximation scheme whereby
corrections in the metric are fed back into the stress tensor.
One then recomputes the corrected metric, feeding these
higher order corrections
back into the stress tensor, and so on. At these higher orders, one
should take into account the quantum fluctuations in the metric
itself, by adding an effective graviton stress tensor to the right
hand side of (1.1). But, to the order we are working here ($O(\hbar)$),
it is reasonable to ignore the quantum metric fluctuations when one
is computing the semiclassical back-reaction caused by the gauge bosons
alone to first order. The consequence of this approach is that the
back-reaction (i.e., the change in the metric) is only meaningful
in a perturbative sense, that is, the \underline{effect} of the gauge
boson stress tensor is regarded as a perturbation of the classical
Schwarzschild geometry. In order to regard $T^{\mu}_{\nu}$ as a perturbation
there are two length scales which are relevant. One is clearly the
mass of the blackhole itself. The other scale comes about because,
as already pointed out earlier, the corrected metric is not asymptotically
flat. Thus, in order that the metric corrections be small compared to
the background metric, one cannot consider the perturbations to be
of unbounded extent. Beyond this cutoff, we can match our solution to
an asymptotically flat spacetime. Naturally, if (1.1) could be exactly
solved, no patching would be expected, and the corrected metric
would presumably be valid for all $r$.

To find this cutoff, we demand that the
magnitude of the ratio of the perturbation to the background
metric be small:
$$ |{{\delta g} \over {\hat g}}| \equiv \delta < 1.\eqno(2.23)$$
Subtracting (2.1) from (2.2) yields the components of the metric
perturbation,
$$\delta g_{vv} = 2\epsilon C_2(G)\left[ {M \over r} B(r) -
(1 - {{2M} \over r})
 A(r) \right] ,\eqno(2.24)$$
$$\delta g_{v \tilde r} = \epsilon C_2(G)\, A(r) .\eqno(2.25)$$
It is easy to check that both perturbations have the same asymptotic
magnitude:
$$|{{\delta g} \over {\hat g}}| \rightarrow {{\epsilon C_2} \over {6K}}
\, ({r \over M})^2.$$
Inserting this limiting form
into (2.23) we find
$$1  \leq ({r_{dom} \over {2M}})^2 = {{3K} \over {2 C_2(G)}}   \,
({{\delta} \over {\epsilon}}).
\eqno(2.26)$$
The domain radius $r_{dom}$ is defined to be the upper bound for which
the metric ratios in (2.23) are uniformly small, i.e.,
$|\delta g| < |g|$ for $2M < r < r_{dom}$.
To get a feeling for the domain size, consider the case where
$\delta = \epsilon$. Solving for $r_{dom}/M$ from (2.26) shows that
the domain radius ranges between
roughly $8M$ and $27M$ (Table I) and we see
that all the antigravity regions fall comfortably within
their respective domains for $C_2 \stackrel{<}{\sim} 1000$.
This gives us confidence that these solutions
represent fairly reasonable approximations to the
back-reaction problem computed to this order.
For a given $\epsilon$, as $C_2(G)$ increases, the outer radius
$r_2$ increases but the radius of the domain of validity decreases.
Hence, there must exist a maximal group for which the
perturbative calculation
breaks down. This breakdown occurs when $r_2  \approx r_{dom}$.
This coincidence limit is achieved for groups with
$C_2(G) \sim 1000$, and the corresponding radius limit is about
$r_2 \sim 8M$ (Table I).
Of course, beyond the cutoff radius, we expect an asymptotically
flat spacetime;
one can match the perturbed metric to a Schwarzschild metric
at a radius $r_2 < {\bar r} < r_{dom}$ using the standard
junction conditions [10]; the mass
parameter for the latter geometry
is $m({\bar r}) > 0 $ (see Fig. 3).
That is,
$$- g_{tt}(r > {\bar r})
= 1 - {{2m({\bar r})} \over r},\eqno(2.27)$$
for the $tt$-component of the metric outside the domain of validity.
With this fixed, the $rr$-component will be automatically continuous
across the boundary $r = {\bar r}$ (the angular components are vacuously
continuous).

\noindent
{\bf 3. Discussion}

By solving for the back-reaction of abelian and
non-abelian gauge bosons on a
blackhole, we have shown that the induced bosonic negative energy
density can
produce a net negative mass distribution (positive hole mass plus
negative bosonic mass)
extending beyond the event horizon.
The repulsive force arising from this energy density
can
dominate the attractive pull of the classical blackhole mass
for gauge groups with $C_2(G) \stackrel{>}{\sim} 100$,
and some of the string-inspired
symmetry groups appear to satisfy this criterion. These results are
important because they exhibit, already at lowest order,
the phenomenon of localized gravitational repulsion
or ``defocussing", characteristic of traversable wormholes [4].
Of course, we do not expect a traversable wormhole to form at lowest
order, but the fact that the
requisite localized net negative mass-energy is
manifest suggests that a wormhole,
perhaps dressed by field fluctuations,
could form (maybe as the endpoint of blackhole evaporation)
if the higher
order back-reactions (1.1) drive the complete solution to pinch-off
between the negative mass and ``normal" regions, but leaving some
of the negative mass at the pinch-off site. The following
comparison may make this point clear. The class of traversable
wormholes considered in [12] have the following mass distribution
$$m(r) = m_{shell}\, \delta(r - {\bar r}) + M\,
\Theta(r - {\bar r}),\eqno(3.1)$$
where $m_{shell} < 0$ and $M > 0$. Junction conditions are used to match
two Schwarzschild geometries (of mass $M$) at the
boundary  ${\bar r} > 2M$.
The region containing the blackhole ($r < {\bar r}$)
is excised by hand; thus the
wormhole throat is located at ${\bar r}$, precisely where the
total net mass goes
negative (note, by the way, the derivative $(dz/dr)$ is finite
near the throat
but discontinuous at $r = {\bar r}$,
unlike the canonical Schwarzschild case,
where it diverges on the throat).
A glance back at Fig. 2 reveals a qualitative similarity between
(3.1) and the mass function derived here. As mentioned earlier, we also
match on a Schwarzschild geometry at for
$r_2 < {\bar r} < r_{dom}$ (of mass
$ m({\bar r}) > 0 $, see Fig. 3). Thus, our mass distribution (2.17)
also will also include a step
function term, $m({\bar r})\,\Theta(r - {\bar r})$,
and the negative mass ``well" in Fig. 2. is qualitatively similar
to the the first term in (3.1).

An analysis of the domain of validity
of our calculations shows the location of the
repulsive region
lies within this domain for a range of the
perturbation parameters. This is comforting and suggests the
trend revealed by these calculations
may persist at higher orders.
The next iteration step in the back-reaction will be sensitive
to the negative energy density of the gauge bosons and should
be important when the
magnitude of the integrated mass located between
$r_1$ and $r_2$, $\int_{r_1}^{r_2}
\mu(r) d^3r$, is of order $M$.

The limit $\epsilon \rightarrow 1$ corresponds to the Planck mass
limit $M \rightarrow M_{pl}$. We see from Table I that the radii
$r_2$
increase, for a fixed $C_2$, in this limit. The blackhole will
of course evaporate, but the endpoint of this evaporation
could lead
to a wormhole (for a large enough $C_2$).
Although we
have not taken into account the effect of quantum metric
fluctuations, it would be interesting to include their
effect
by introducing a renormalized graviton stress tensor on the
right hand side of (1.1). Such a tensor has in fact been calculated
in an arbitrary static vacuum background, but requires ``tailoring" before
it can be incorporated in the present calculation [13]. If we
are merely interested in the sign of the graviton's renormalized
energy density, a shortcut and clue are provided by the
fact that primordial gravitons are ``squeezed'' by cosmological
expansion [14].
Applying the analysis of [6] implies they
will have a {\it negative} energy density, and so we can hope
that the graviton contribution to (1.1) enhances
the repulsive effect of the gauge bosons.

Conformally coupled ($\xi = 1/6$) massless
scalars also contribute to the
dilation effect, but the depth of the corresponding pocket
of negative energy is roughly one hundred and twenty
times smaller than the vector boson
case and the range is down by a factor of three [15]
(The back-reaction of conformally coupled scalars on a
blackhole has been
treated by York [8]).
To explore
the scalar back-reaction case completely, we need to know the
scalar stress tensor
in a Schwarzschild background for arbitrary coupling $\xi$, where
$\xi \phi^2 R$ is the non-minimal coupling of scalars to
gravity.
To our knowledge, this tensor has not yet been calculated.
Scalar feedback becomes particularly interesting when
looking for wormholes in certain classes of higher-derivative
gravity theories [16]. In those models, the higher powers of the
curvature are mapped onto an equivalent scalar sector and the scalar
couples {\it minimally} (i.e, $\xi = 0$) to the metric.

\vspace{1cm}
\noindent
{\bf Acknowledgment}
We are grateful to James W. York, Jr. for carefully reading a preliminary
version of the manuscript and for valuable comments
that helped us correctly interpret our results.
This work was supported in part by DOE Grant No. DE-FGO5-
85ER40226.

\vfill\eject

\vspace{1cm}
\noindent
{\bf References}

 \noindent
[1]. See, e.g., S. Weinberg, Rev. Mod. Phys. {\bf 61}, 1 (1989).

\noindent
[2]. L. Flamm, Physik Z. {\bf 17}, 448 (1916).

\noindent
[3]. M.D. Kruskal, Phys. Rev. {\bf 119}, 1743 (1960).

\noindent
[4]. M.S. Morris and K.S. Thorne, Am. J. Phys. {\bf 56}, 395 (1988).

\noindent
[5]. N.D. Birrell and P.C.W. Davies, {\it Quantum Fields in Curved
Space}, (Cambridge University Press, 1982).

\noindent
[6]. D. Hochberg and T.W. Kephart, Phys. Lett. {\bf B}268, 377 (1991).

\noindent
[7]. B.P. Jensen and A. Ottewill, Phys. Rev. {\bf D}39, 1130 (1989).

\noindent
[8]. J.W. York, Jr., Phys. Rev. {\bf D}31, 775 (1985).

\noindent
[9]. S.W. Hawking and G.F.R. Ellis, {\it The Large Scale Structure
of Space-time}, (Cambridge University Press, 1973), Section 4.3.

\noindent
[10]. C.W. Misner, K.S. Thorne and J.A. Wheeler, {\it Gravitation},
(W.H. Freeman, San Francisco, 1973), Chap. 23.

\noindent
[11]. See, e.g., M.B. Green, J.H. Schwarz and E. Witten,
{\it Superstring Theory},
(Cambridge University Press, 1987), pg. 356;
M. Kaku, {\it Introduction to Superstrings}, (Springer-Verlag,
N.Y., 1988), pg. 402.

\noindent
[12]. M. Visser, Nucl. Phys. {\bf B}328, 203 (1989).

\noindent
[13]. B. Allen, A. Folacci and A.C. Ottewill, Phys. Rev.
{\bf D}38, 1069 (1988).

\noindent
[14]. L.P. Grishchuk and Y.V. Sidorov, Phys. Rev. {\bf D}42, 3413 (1990).

\noindent
[15]. D.N. Page, Phys. Rev. {\bf D}25, 1499 (1982); K.W. Howard and
P. Candelas, Phys. Rev. Lett. {\bf 53}, 403 (1984).

\noindent
[16]. D. Hochberg, Phys. Lett. {\bf B}251, 349 (1990).

\vfill\eject
\mediumtext
\begin{table}
\begin{tabular}{ccccc}
Gauge group: $C_2(G)$ & $\epsilon$,$(\epsilon_{min})$ & $r_1/2M$ &
$ r_2/2M$ & $r_{dom}/2M$ \\ \hline
            &          &          &         &$(\delta =
 \epsilon)$     \\
101 & (1.) & 2.53 & 2.60 & 13.4 \\
120 $(E_8 \times E_8)$  & (.85) & 2.34 & 2.81 & 12.3 \\
    &  1.   & 1.73 & 3.68 &      \\
308 $(E_8 \times [U(1)]^{248})$  & (.34) & 2.38 & 2.75 & 7.7  \\
    &  1.   & 1.15 & 4.78 &      \\
1000 & (.11) & 1.93 & 3.38 & 4.3  \\
     & .2    & 1.30 & 4.60 &      \\
\end{tabular}
\caption{Range of negative mass region and radius of domain of validity vs.
gauge group}
\end{table}
\vspace{3cm}
\begin{center}
FIGURE CAPTIONS
\end{center}

\noindent
Fig. 1. Renormalized energy density for an abelian vector
boson as a function of ${r \over {2M}}$ ($\kappa = {1 \over {4M}}$).
\vspace{.5cm}

\noindent
Fig. 2. The effective bosonic mass-function.
\vspace{.5cm}

\noindent
Fig. 3. Derivative-squared of the embedding function. The
dotted line represents the negative
mass region between $r_1$ and $r_2$ (for $\epsilon \approx .3$ and
$C_2 \approx 300$).
An asymptotically flat geometry ($\epsilon = 0$) can be joined
on at a
cutoff radius $r_2 < {\bar r}  < r_{dom}$.

\end{document}